\documentclass[%
 reprint,
 superscriptaddress,
showpacs,preprintnumbers,
 amsmath,amssymb,
 aps,
 prc,
floatfix,
]{revtex4-1}

\usepackage{graphicx}% Include figure files
\usepackage{dcolumn}% Align table columns on decimal point
\usepackage{bm}% bold math
\usepackage{mathtools}
\usepackage{amsmath}
\usepackage{gensymb}
\usepackage{hyperref}
\usepackage{lineno}
\usepackage{xcolor}
\begin{document}

\preprint{LA-UR-22-30616}

\title{$\beta^{+}$ Gamow-Teller strengths from unstable $^{14}$O via the $(d,{}^2\text{He})$ reaction in inverse kinematics}

\author{S. Giraud}
\email{giraud@nscl.msu.edu}
\affiliation{National Superconducting Cyclotron Laboratory, Michigan State University, East Lansing, MI 48824, USA\looseness=-1}
\affiliation{Joint Institute for Nuclear Astrophysics: Center for the Evolution of the Elements, Michigan State University, East Lansing, MI 48824, USA\looseness=-1}
\affiliation{Department of Physics and Astronomy, Michigan State University, East Lansing, MI 48824, USA\looseness=-1}

\author{J. C. Zamora}
%\email{zamora@nscl.msu.edu}
\affiliation{National Superconducting Cyclotron Laboratory, Michigan State University, East Lansing, MI 48824, USA\looseness=-1}

\author{R. Zegers}
\email{zegers@nscl.msu.edu}
\affiliation{National Superconducting Cyclotron Laboratory, Michigan State University, East Lansing, MI 48824, USA\looseness=-1}
\affiliation{Joint Institute for Nuclear Astrophysics: Center for the Evolution of the Elements, Michigan State University, East Lansing, MI 48824, USA\looseness=-1}
\affiliation{Department of Physics and Astronomy, Michigan State University, East Lansing, MI 48824, USA\looseness=-1}

\author{D. Bazin}
\affiliation{National Superconducting Cyclotron Laboratory, Michigan State University, East Lansing, MI 48824, USA\looseness=-1}
\affiliation{Department of Physics and Astronomy, Michigan State University, East Lansing, MI 48824, USA\looseness=-1}

\author{Y. Ayyad}
\affiliation{IGFAE, Universidade de Santiago de Compostela, E-15782 Santiago de Compostela, Spain\looseness=-1}
\affiliation{National Superconducting Cyclotron Laboratory, Michigan State University, East Lansing, MI 48824, USA\looseness=-1}

\author{S. Bacca} 
\affiliation{Institut f\"ur Kernphysik and PRISMA$^+$ Cluster of Excellence, Johannes Gutenberg-Universit\"at, 55128 Mainz, Germany}
\affiliation{Helmholtz-Institut Mainz, Johannes Gutenberg-Universit\"at Mainz, D-55099 Mainz, Germany\looseness=-1}

\author{S. Beceiro-Novo}
\affiliation{Department of Physics and Astronomy, Michigan State University, East Lansing, MI 48824, USA\looseness=-1}

\author{B.A. Brown}
% \email[]{}
\affiliation{National Superconducting Cyclotron Laboratory, Michigan State University, East Lansing, MI 48824, USA\looseness=-1}
\affiliation{Joint Institute for Nuclear Astrophysics: Center for the Evolution of the Elements, Michigan State University, East Lansing, MI 48824, USA\looseness=-1}
\affiliation{Department of Physics and Astronomy, Michigan State University, East Lansing, MI 48824, USA\looseness=-1}

\author{A. Carls}
\affiliation{National Superconducting Cyclotron Laboratory, Michigan State University, East Lansing, MI 48824, USA\looseness=-1}
\affiliation{Joint Institute for Nuclear Astrophysics: Center for the Evolution of the Elements, Michigan State University, East Lansing, MI 48824, USA\looseness=-1}
\affiliation{Department of Physics and Astronomy, Michigan State University, East Lansing, MI 48824, USA\looseness=-1}

\author{J. Chen}
\affiliation{National Superconducting Cyclotron Laboratory, Michigan State University, East Lansing, MI 48824, USA\looseness=-1}
\affiliation{Physics Division, Argonne National Laboratory, Lemont, IL, 60439, USA\looseness=-1}

\author{M. Cortesi}
\affiliation{National Superconducting Cyclotron Laboratory, Michigan State University, East Lansing, MI 48824, USA\looseness=-1}

\author{M. DeNudt}
\affiliation{National Superconducting Cyclotron Laboratory, Michigan State University, East Lansing, MI 48824, USA\looseness=-1}
\affiliation{Joint Institute for Nuclear Astrophysics: Center for the Evolution of the Elements, Michigan State University, East Lansing, MI 48824, USA\looseness=-1}
\affiliation{Department of Physics and Astronomy, Michigan State University, East Lansing, MI 48824, USA\looseness=-1}

\author{G. Hagen}
\affiliation{Physics Division, Oak Ridge National Laboratory, Oak Ridge, Tennessee 37831, USA\looseness=-1}
\affiliation{Department of Physics and Astronomy, University of Tennessee, Knoxville, Tennessee 37996, USA\looseness=-1}

\author{C. Hultquist}
\affiliation{National Superconducting Cyclotron Laboratory, Michigan State University, East Lansing, MI 48824, USA\looseness=-1}
\affiliation{Joint Institute for Nuclear Astrophysics: Center for the Evolution of the Elements, Michigan State University, East Lansing, MI 48824, USA\looseness=-1}
\affiliation{Department of Physics and Astronomy, Michigan State University, East Lansing, MI 48824, USA\looseness=-1}

\author{C. Maher}
\affiliation{National Superconducting Cyclotron Laboratory, Michigan State University, East Lansing, MI 48824, USA\looseness=-1}
\affiliation{Joint Institute for Nuclear Astrophysics: Center for the Evolution of the Elements, Michigan State University, East Lansing, MI 48824, USA\looseness=-1}
\affiliation{Department of Physics and Astronomy, Michigan State University, East Lansing, MI 48824, USA\looseness=-1}

\author{W. Mittig}
\affiliation{National Superconducting Cyclotron Laboratory, Michigan State University, East Lansing, MI 48824, USA\looseness=-1}
\affiliation{Department of Physics and Astronomy, Michigan State University, East Lansing, MI 48824, USA\looseness=-1}

\author{F. Ndayisabye}
\affiliation{National Superconducting Cyclotron Laboratory, Michigan State University, East Lansing, MI 48824, USA\looseness=-1}
\affiliation{Joint Institute for Nuclear Astrophysics: Center for the Evolution of the Elements, Michigan State University, East Lansing, MI 48824, USA\looseness=-1}
\affiliation{Department of Physics and Astronomy, Michigan State University, East Lansing, MI 48824, USA\looseness=-1}

\author{S. Noji}
\affiliation{National Superconducting Cyclotron Laboratory, Michigan State University, East Lansing, MI 48824, USA\looseness=-1}

\author{S. J. Novario}
\affiliation{Physics Division, Oak Ridge National Laboratory, Oak Ridge, Tennessee 37831, USA\looseness=-1}
\affiliation{Department of Physics and Astronomy, University of Tennessee, Knoxville, Tennessee 37996, USA\looseness=-1}

\author{J. Pereira}
\affiliation{National Superconducting Cyclotron Laboratory, Michigan State University, East Lansing, MI 48824, USA\looseness=-1}
\affiliation{Joint Institute for Nuclear Astrophysics: Center for the Evolution of the Elements, Michigan State University, East Lansing, MI 48824, USA\looseness=-1}

\author{Z. Rahman}
\affiliation{National Superconducting Cyclotron Laboratory, Michigan State University, East Lansing, MI 48824, USA\looseness=-1}
\affiliation{Joint Institute for Nuclear Astrophysics: Center for the Evolution of the Elements, Michigan State University, East Lansing, MI 48824, USA\looseness=-1}
\affiliation{Department of Physics and Astronomy, Michigan State University, East Lansing, MI 48824, USA\looseness=-1}

\author{J. Schmitt}
\affiliation{National Superconducting Cyclotron Laboratory, Michigan State University, East Lansing, MI 48824, USA\looseness=-1}
\affiliation{Joint Institute for Nuclear Astrophysics: Center for the Evolution of the Elements, Michigan State University, East Lansing, MI 48824, USA\looseness=-1}
\affiliation{Department of Physics and Astronomy, Michigan State University, East Lansing, MI 48824, USA\looseness=-1}

\author{M. Serikow}
\affiliation{National Superconducting Cyclotron Laboratory, Michigan State University, East Lansing, MI 48824, USA\looseness=-1}
\affiliation{Joint Institute for Nuclear Astrophysics: Center for the Evolution of the Elements, Michigan State University, East Lansing, MI 48824, USA\looseness=-1}
\affiliation{Department of Physics and Astronomy, Michigan State University, East Lansing, MI 48824, USA\looseness=-1}

\author{L. J. Sun}
\affiliation{National Superconducting Cyclotron Laboratory, Michigan State University, East Lansing, MI 48824, USA\looseness=-1}
\affiliation{Joint Institute for Nuclear Astrophysics: Center for the Evolution of the Elements, Michigan State University, East Lansing, MI 48824, USA\looseness=-1}

\author{J. Surbrook}
\affiliation{National Superconducting Cyclotron Laboratory, Michigan State University, East Lansing, MI 48824, USA\looseness=-1}
\affiliation{Joint Institute for Nuclear Astrophysics: Center for the Evolution of the Elements, Michigan State University, East Lansing, MI 48824, USA\looseness=-1}
\affiliation{Department of Physics and Astronomy, Michigan State University, East Lansing, MI 48824, USA\looseness=-1}

\author{N. Watwood}
\affiliation{National Superconducting Cyclotron Laboratory, Michigan State University, East Lansing, MI 48824, USA\looseness=-1}
\affiliation{Joint Institute for Nuclear Astrophysics: Center for the Evolution of the Elements, Michigan State University, East Lansing, MI 48824, USA\looseness=-1}
\affiliation{Department of Physics and Astronomy, Michigan State University, East Lansing, MI 48824, USA\looseness=-1}

\author{T. Wheeler}
\affiliation{National Superconducting Cyclotron Laboratory, Michigan State University, East Lansing, MI 48824, USA\looseness=-1}
\affiliation{Joint Institute for Nuclear Astrophysics: Center for the Evolution of the Elements, Michigan State University, East Lansing, MI 48824, USA\looseness=-1}
\affiliation{Department of Physics and Astronomy, Michigan State University, East Lansing, MI 48824, USA\looseness=-1}

\date{\today}% It is always \today, today,
             %  but any date may be explicitly specified

\begin{abstract}
For the first time, the $(d,{}^2\text{He})$ reaction was successfully used in inverse kinematics to extract the Gamow-Teller transition strength in the $\beta^{+}$ direction from an unstable nucleus. The nucleus studied was $^{14}$O, and the Gamow-Teller transition strength to $^{14}$N was extracted up to an excitation energy of 22 MeV. The measurement of the $(d,{}^2\text{He})$ reaction in inverse kinematics was made possible by the combination of an active target time projection chamber and a magnetic spectrometer. The data were used to test shell-model and state-of-the-art coupled cluster calculations. Shell-model calculations reproduce the measured Gamow-Teller strength distribution up to about 15 MeV reasonably well, after the application of a phenomenological quenching factor. Coupled-cluster calculation reproduces the full strength distribution well without such quenching, owing to the large model space, the inclusion of strong correlations, and the coupling of the weak interaction to two nucleons through two-body currents. This indicates that such calculations provide a very promising path for answering long-standing questions about the observed quenching of Gamow-Teller strengths in nuclei.
\end{abstract}

%\pacs{Valid PACS appear here}
\maketitle

%\begin{linenumbers}
Nuclear charge-exchange (CE) reactions provide important tools for studying isovector excitations and studying the spin-isospin response of nuclei. They provide important information about nuclear structure, bulk properties of nuclei, and processes mediated by the weak nuclear force \cite{Osterfeld1992,Rapaport1994,HAR01, Ichimura2006,RevModPhys.75.819,Fujita2011549,Frekers2018,Langanke2021}. Of particularly high impact has been the ability to indirectly and model-independently extract Gamow-Teller (GT) transitions strength from CE experiments at intermediate beam energies ($E \gtrsim 100$ MeV/nucleon). GT transitions are associated with the transfer of spin ($\Delta S=1$), isospin ($\Delta T=1$), and no angular momentum ($\Delta L=0$), and mediate allowed $\beta$ decay and electron-captures (ECs). The latter play important roles in astrophysical phenomena \cite{RevModPhys.75.819,Langanke2021}, such as core-collapse supernovae \cite{Bethe.Brown.ea:1979,Fuller.Fowler.Newman:1982a,Janka.Langanke.ea:2007}, thermonuclear supernovae \cite{Brachwitz.Dean.ea:2000, Iwamoto.Brachwitz.ea:1999}, and the crusts of neutron stars that accrete material from binary-system companions \cite{Schatz.Gupta.ea:2014,Haensel.Zdunik:1990}. 

The extraction of the GT transition strengths ($B$(GT)) from CE reactions, which are mediated by the strong nuclear force, is possible because of the well-established proportionality between the extracted differential cross section at small momentum transfer ($q\approx0$) and $B$(GT) \cite{Taddeucci1987}, valid for $B$(GT)$\gtrsim0.01$ \cite{Zegers:2006}. The proportionality factor, referred to as the unit cross section ($\hat{\sigma}_{\textsc{GT}}$), is conveniently calibrated using transitions for which the $B$(GT) is known directly from $\beta$/EC decay. Crucially, unlike $\beta$/EC decay, the extraction of $B$(GT) from CE experiments is not limited to a finite $Q$-value window, which is particularly important for the above-mentioned astrophysical applications, where, due to high stellar temperatures and/or densities, EC transitions to highly excited states can occur. For constraining the astrophysical EC rates, charge-exchange reactions in the ($n$,$p$) direction are key, but over the past four decades, many CE experiments in both the ($p$,$n$) and ($n$,$p$) direction using different probes have been performed on targets of stable nuclei aimed at elucidating the structure of nuclei, determining astrophysical reaction rates, and constraining neutrino-induced reactions and (double) $\beta$-decay rates. 

In the astrophysical scenarios mentioned above, ECs on many unstable isotopes play crucial roles. Compared to ($p$,$n$) CE reactions, for which experiments in inverse kinematics (i.e., with the unstable nucleus of interest being produced as the beam) were successfully performed to study GT transitions in the $\beta^{-}$ direction up to high excitation energy and across the chart of isotopes \cite{COR96, BRO96,SHI97,TER97,TAK01,LI02,Sasano.Perdikakis.ea:2011,PhysRevLett.121.132501,KOB15,LIP18}, the development of CE experiments in the ($n$,$p$) direction in inverse kinematics with beams of unstable isotopes has been more challenging. The $(^7\text{Li},{}^7\text{Be}+\gamma)$ reaction was successfully developed to study ($n$,$p$)-type CE reactions in inverse kinematics \cite{Zegers2010,PhysRevLett.108.122501}, but it can only be used for light ($A\lesssim 35$) nuclei and for excitation energies below particle-separation energies. In this Letter, we present the successful development of an alternative method, namely the $(d,{}^2\text{He})$ reaction in inverse kinematics, which can be used to extract $B$(GT) up to high excitation energies and without intrinsic limitations on the mass of the isotope.

In the first experiment presented here, Gamow-Teller transitions from unstable $^{14}$O to $^{14}$N were studied to extract the GT strength distribution up to excitation energies of $\sim 20$ MeV. 
The measurement complements previous studies in the $A=14$ multiplet \cite{Wharton1980,Anantaraman1986,PhysRevLett.97.062502,AJZENBERGSELOVE19911,Goodman1980,Taddeucci1984,Rapaport1987,Hernandez1981}. The $A=14$ nuclei have posed significant challenges to theoretical calculations and require the inclusion of three-body forces and an accurate treatment of two-body currents, for example to explain the anomalously long half-lives (corresponding to very small GT transition strengths)  for the analog $\beta$ decays from $^{14}$O and $^{14}$C to ground state (g.s.) of $^{14}$N \cite{PhysRevLett.106.202502, PhysRevLett.100.062501, PhysRevC.79.054331, PhysRevLett.113.262504}. In recent effective field theory (EFT) calculations \cite{Gysbers2019}, it is possible to explain, based on first principles, the reduction of GT strength observed in experimental data compared to other theoretical calculations, including the shell model (SM). This reduction, referred to as the ``quenching'' of the GT strength \cite{TOWNER1987263,GAARDE1981258,BRO88} has important implications for the astrophysical applications, such as the ones mentioned above, as well as fundamental phenomena such as neutrinoless double-$\beta$ decay \cite{ENG17,SUH17}. By including strong correlations present in the nucleus and the coupling of the weak interaction to two nucleons through two-body currents (2BCs) in addition to the one-body Gamow-Teller operator, it is possible to describe the quenching in the EFT calculations \cite{Gysbers2019,PhysRevC.102.025501}. It is important to further test such calculations by comparing GT transition strength up to high excitation energies, including for nuclei far from stability, and charge-exchange reactions in inverse kinematics are excellent tools for this purpose. Through the development of the $(d,{}^2\text{He})$ reaction in inverse kinematics, we demonstrate that experimentally viable methods now exist in both the $\beta^{+}$/EC and $\beta^{-}$ directions to carry out such studies.

The $(d,{}^2\text{He})$ reaction in forward kinematics, i.e. with a deuteron beam, has been used to study many stable nuclei~\cite{PhysRevC.47.648, Okamura1995, PhysRevC.52.R1161, Frekers2004}. The $(d,{}^2\text{He})$ reaction refers to a $(d,2p)$ reaction for which the two outgoing protons couple to a $^1$S$_0$ ($T=1$) state. This state is unbound by $\approx0.5$~MeV. Contributions from higher partial waves become significant at higher internal energy $\epsilon _{pp}\gtrsim 4$~MeV~\cite{Kox1993}. The wave function of the deuteron is dominated by the $^3$S$_1$ ($T=0$) configuration. Therefore, the $(d,{}^2\text{He})$ reaction at low $^2$He internal energy ($\epsilon _{pp} \lesssim 2$~MeV) is selective to excitations involving the transfer of spin ($\Delta S=1$)~\cite{Kox1993, PhysRevC.47.648}. In inverse kinematics, and for $(d,{}^2\text{He})$ reactions at $q\approx 0$, the energy of the two protons emitted is very small, making the use of a foil (e.g, CD$_2$) or liquid deuterium unfeasible. Therefore, a gaseous active-target time-projection chamber (AT-TPC)~\cite{Bradt2017} was used in the present work, in which the deuterium gas served as both the target and the tracking medium for the protons. The beam-like fragment (i.e., $^{14}$N, or one of its decay products) was detected in the S800 spectrograph~\cite{Bazin2003} to serve as a trigger for CE events.

A 10- to 50-pnA, 150-MeV/nucleon beam of $^{16}$O was accelerated by the Coupled Cyclotron Facility at the National Superconducting Cyclotron Laboratory (NSCL), and struck a 1316 mg/cm$^2$-thick Be production target. A 150 mg/cm$^2$-thick Al degrader was used in the A1900 fragment separator~\cite{Morrissey2003} to produce a 70\%-pure $^{14}$O beam with intensities between 0.2 to 0.7 Mpps. The time-of-flight (TOF) between two scintillators placed at the exit of the A1900 and the entrance of the S800 Spectrograph~\cite{Bazin2003} beam line (the S800 object point) was used to separate $^{14}$O from $^{13}$N (23\%) and $^{12}$C (7\%) contaminants on an event-by-event basis. The AT-TPC~\cite{Bradt2017}, used for the first time with a fast rare-isotope beam, was filled with pure deuterium gas at a pressure of 530 Torr ($\pm$0.5\%), corresponding to a thickness of 11.7 mg/cm$^2$. The gas of the active volume was isolated from the beam-line and S800 vacuum by 12 $\mu$m-thick polyamide windows. In the AT-TPC, a 500-V/cm uniform electric field, directed along the beam axis, drifts electrons produced by ionizing charged particles upstream towards a micromegas pad plane with 10240 independent readout channels, which provides the transverse track images.  The third position coordinate, along the beam direction, is determined from the drift time of the electrons. The drift velocity was $\approx$ 0.9 cm/$\mathrm \mu s$. The pad plane has a central aperture of 3 cm diameter to allow the beam to enter the AT-TPC. Hence, tracks from the beam particles and outgoing beam-like fragments are not observed in the AT-TPC. The fragment identification was performed event-by-event using the TOF between scintillators at the S800 object point and at the focal plane of S800, and the energy loss of the fragments in an ionization chamber at the focal plane of S800. The object scintillator was also used to monitor the beam rate for the cross-section determination and the focal plane scintillator was used to trigger a $\approx$110 $\mu$s long readout window for signals from the AT-TPC. Momenta of the fragments at the target were reconstructed from the positions and angles measured with two cathode-readout drift chambers~\cite{Yurkon1999} at the S800 focal plane. These reconstructed quantities were used for S800 acceptance corrections and determining absolute cross-sections.\\

The excitation energy of $^{14}$N produced after the $^{14}$O$(d,{}^2\text{He})$ reaction is reconstructed in three steps. First, the electron cloud produced by ionizing charged particles in the AT-TPC is analyzed with pattern recognition and fitting routines to identify the $(d,{}^2\text{He})$ events and extract the $^2$He momentum \cite{Ayyad2018,Zamora2021}. The selection of $(d,{}^2\text{He})$ events is ensured by the coincidence between an identified $^{14}$O ion in the beam, a relevant residual fragment in the S800, and two fitted tracks with a minimal distance smaller than 2$\sigma$ of the minimal-distance distribution ($\lesssim 1$~cm) in the beam region. The point of minimal distance defines the reaction vertex. The energy (deduced from the range) and angular resolutions of a single track are about 15 keV and $1.5\degree$, respectively. The momentum vectors of the two protons are used to reconstruct the $^{2}$He momentum vector and, through an invariant-mass calculation, its internal energy. Finally, the excitation energy of $^{14}$N is obtained from a missing-mass calculation. 
\\
\begin{figure}
\hspace*{-1.2cm}
\includegraphics[width=0.61\textwidth,clip]{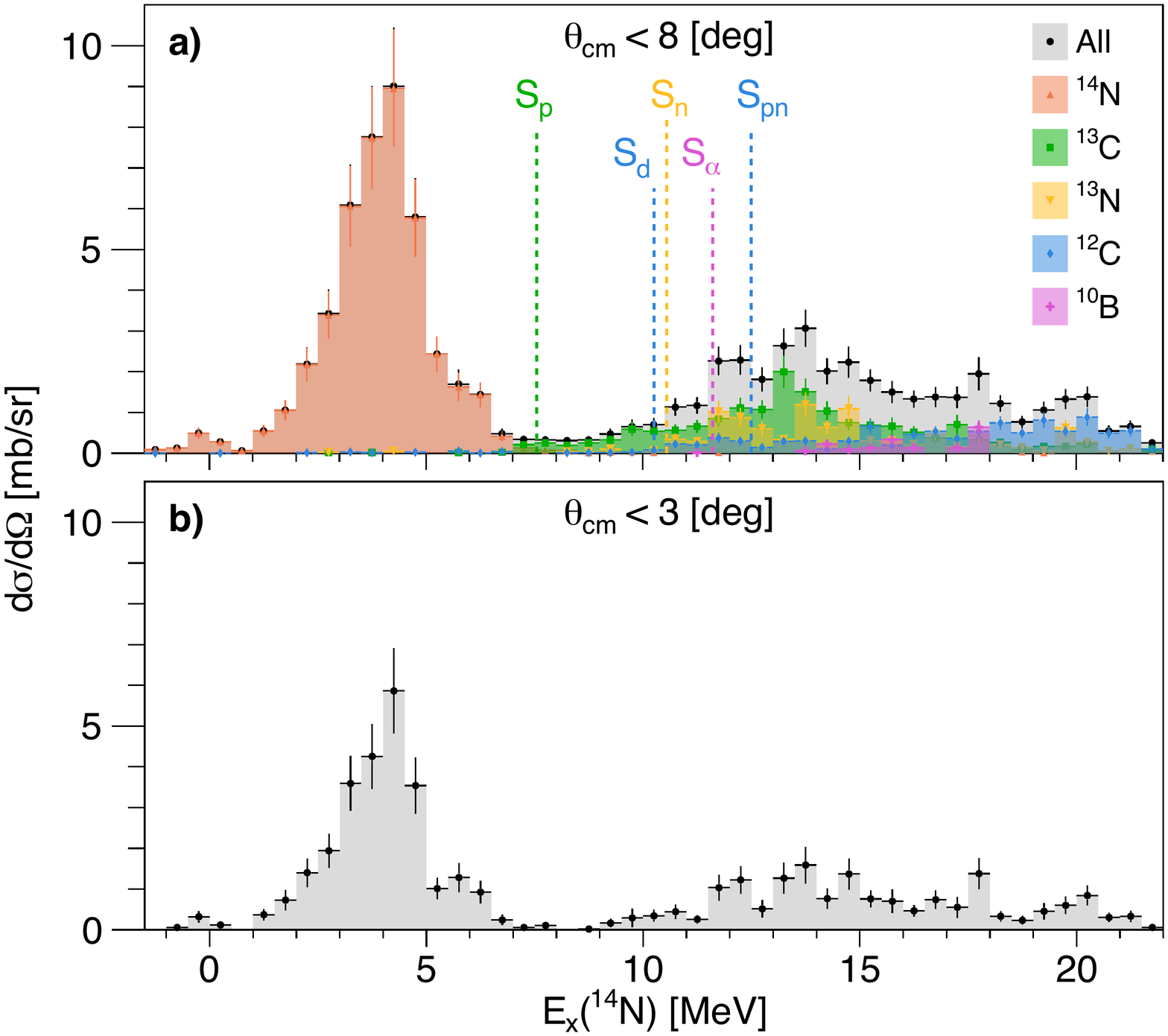}
\caption{Differential cross sections for the $^{14}$O$(d,{}^2\text{He})$ reaction a) for the entire scattering angle range ($\theta_\text{c.m.} < 8 \degree$) and b) for $\theta_\text{c.m.} < 3\degree$. In a) the colors represent the contributions from the different decay channels corresponding to respective fragments detected in the S800 spectrometer, and the dashed lines indicates the $^{14}$N particle-decay thresholds ($S_p$ = 7.55 MeV, $S_n$ = 10.55 MeV, $S_d$ = 10.26 MeV, $S_{\alpha}$ = 11.61 MeV and $S_{pn}$ = 12.50 MeV).}\label{fig:CS_Ex}
\end{figure}

Simulations for estimating the efficiency and acceptance of the experiment were performed with the \textsc{attpcroot} package \cite{Ayyad_2017,ATTPCROOT}, using a $(d,{}^2\text{He})$ event generator based on calculated cross-sections using the code Adiabatic Coupled-Channels Born Approximation \textsc{accba}~\cite{Okamura1995,Okamura1999} specifically developed for $(d,{}^2\text{He})$ reactions. The code reproduces well the differential cross sections for $(d,{}^2\text{He})$ reactions performed in forward kinematics~\cite{Rakers2002,Grewe2004}. For the entrance and exit channels, optical-model parameters obtained from the Koning-Delaroche phenomenological potential~\cite{Koning2003} for protons and deuterons (extended parameterization in the code \textsc{talys}~\cite{Koning2007,Koning2012}) were used. The Love-Franey effective nucleon-nucleon interaction at 140 MeV~\cite{Franey1985} was used. The spectroscopic amplitudes of the transitions were obtained from SM calculations in the $p$-shell model space and the CKII~\cite{Cohen1965} interaction using the \textsc{nushellx}~\cite{Brown2014} code. A strong asset of \textsc{attpcroot} simulation is that the analog signal of each pad of the sensor plane is analyzed in the same manner as the experimental data. In addition, the reconstruction of the simulated events is performed with the same algorithms as for the experimental data. Therefore, the simulation also provides realistic estimations of the efficiency and acceptance of the AT-TPC.

Figure~\ref{fig:CS_Ex}(a) shows the extracted differential cross sections as a function of the excitation energy of $^{14}$N at scattering angles below 8$\degree$, gated on $^{14}$N, or its decay products ($^{13}$C, $^{13}$N, $^{12}$C and $^{10}$B) identified with the S800 for the excitation energies above the $^{14}$N particle-decay thresholds, as indicated in the figure. The experimental spectra are almost background free due to the availability of a set of stringent conditions for selecting $(d,{}^2\text{He})$ events. Please note that these spectra are integrated over the $\epsilon_{pp}$ range accepted by the AT-TPC. At small scattering angles and small reaction $Q$ value, the two protons have the lowest energies, requiring $\epsilon_{pp}$ to be within 1.5 and 2.5 MeV for both protons to escape the central insensitive region and to have paths that end within the chamber. At larger scattering angles and $Q$ values, only events with $\epsilon_{pp}<1$ MeV have path lengths that end inside the AT-TPC and can be reconstructed. Figure~\ref{fig:CS_Ex}(b) shows the differential cross sections for scattering angles below 3$\degree$, integrated for all the decay products. Near $Q=0$ ($E_{x}=3.7$ MeV), the excitation energy is almost completely determined by the angle of the reconstructed $^{2}$He particle, and the resolution is limited to $\sim 2.1$ MeV full-width at half-maximum value. At smaller and larger $Q$ values, the energy of the $^{2}$He particle is also important, and a resolution of $\sim 1.2$ MeV can be achieved.

\begin{figure}
\hspace*{-0.4cm}
\includegraphics[width=0.5\textwidth,clip]{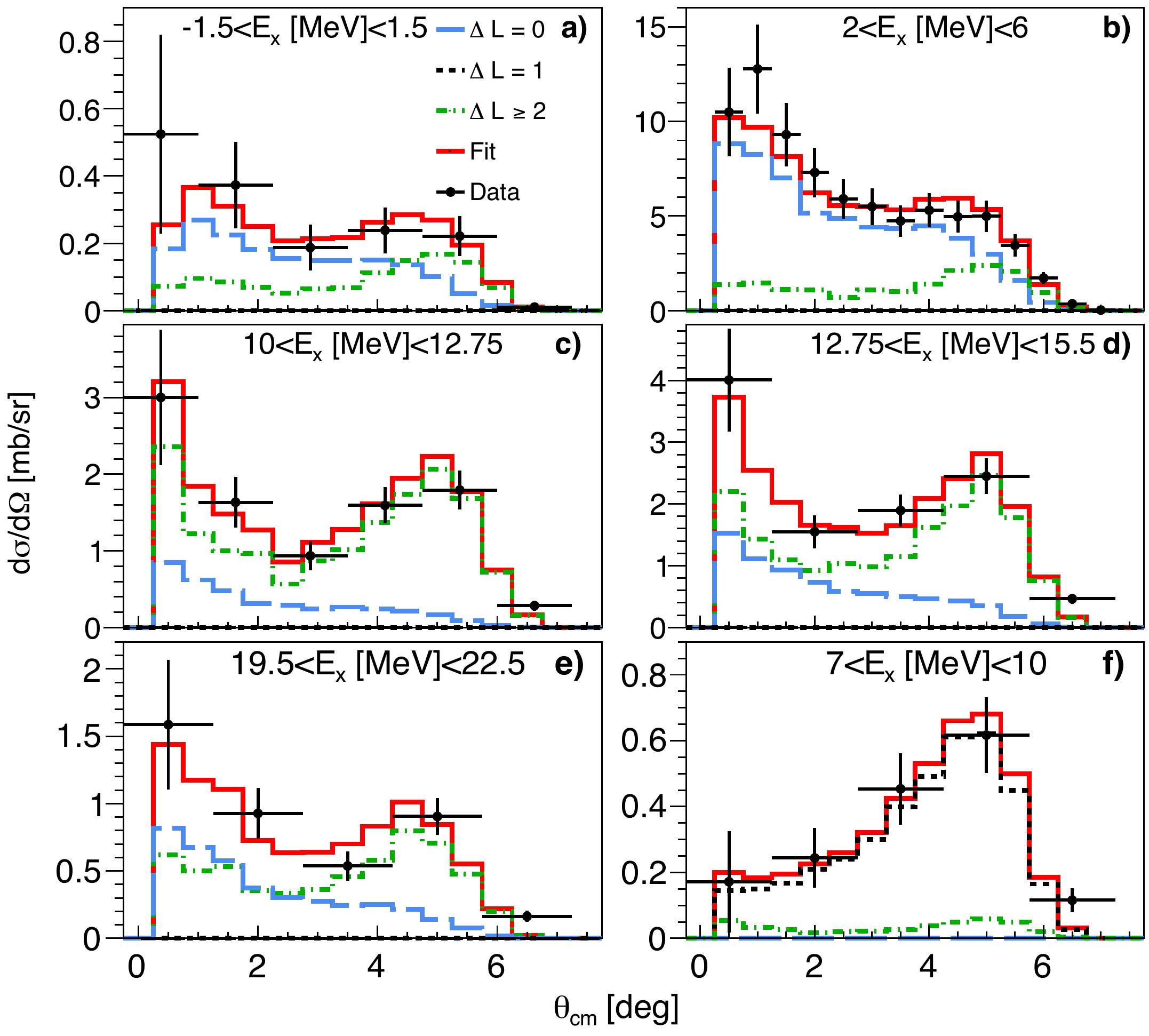}
\caption{Angular distribution for different ranges in energy with MDA results, using angular distributions from the \textsc{attpcroot} simulation. The error bars represent statistical and systematic errors. a) to e) are gated on excitation energy ranges with peaks, see Fig.~\ref{fig:CS_Ex}, and f) is gated on a region without strong GT transition.}\label{fig:MDA}
\end{figure}

Figure \ref{fig:MDA} shows the differential cross sections for several key excitation-energy regions. The error bars in the data include statistical and systematic uncertainties. The latter are dominated by uncertainties in the acceptance corrections and beam-intensity determination, but are relatively small compared to the statistical uncertainties, except for the state at 3.95 MeV for which both types of uncertainty are comparable. In order to extract the $\Delta L=0$ (GT) contributions from the excitation-energy spectra, a multipole decomposition analysis (MDA)~\cite{Bonin1984, MOINESTER198941} was performed. For each region, the experimental differential cross sections were fitted with a linear combination of calculated angular distributions associated with angular momentum transfers $\Delta L=0$, 1, and 2. The lower the angular momentum transfers are, the more forward-peaked the differential cross sections are. Transitions with $\Delta L>2$ are suppressed near $q=0$ and are not included in the fit, but minor contributions might be absorbed in the extracted contributions from components with $\Delta L=1$ and 2, as their angular distributions all peak at larger scattering angles. Prior to the fit, the calculated differential cross sections for each $\Delta L$ from \textsc{accba} were inserted in the \textsc{attpcroot} simulation to account for the $\epsilon_{pp}$ acceptance of the AT-TPC as a function of scattering angle and $Q$-value, as discussed above. As a consequence, the cross sections at larger angles are somewhat enhanced compared to a calculation for fixed $\epsilon_{pp}$.
The MDA results are shown in Fig.~\ref{fig:MDA} as colored histograms.   

The dominant peak in the spectra of Fig.~\ref{fig:CS_Ex} is the transition to the $1^{+}$ state at 3.95 MeV, which has a known $B$(GT) of 2.73 from $\beta$ decay~\cite{Hernandez1981}. As expected for a strong GT transition, it is dominated by $\Delta L=0$, as shown in Fig.~\ref{fig:MDA}(b). The $0^{+}$ isobaric analog of the $^{14}$O g.s. state at 2.31 MeV is not notably excited, as expected for the $\Delta S=1$ $(d,{}^2\text{He})$ reaction, unlike the case for the isospin-symmetric $^{14}$C($p$,$n$) reaction ~\cite{Goodman1980,Taddeucci1984,Rapaport1987}.  

To compare the unit cross section~\cite{Taddeucci1987} from the present experiment with that obtained from previous $(d,{}^2\text{He})$ experiments in forward kinematics~\cite{Rakers2002}, the $\Delta L=0$ fraction of the cross section obtained after MDA was extrapolated to $\epsilon _{pp}<1$ MeV and $q=0$ on the basis of the \textsc{accba} calculations. A unit cross section of $\hat{\sigma}_{\textsc{GT}}=2.74 \pm 0.29$~mb/sr was obtained, in good agreement with the value of  $2.58 \pm 0.14$~mb/sr found for the $^{12}$C(d,$^2$He) reaction in forward kinematics~\cite{Rakers2002} at a beam energy of 85 MeV/nucleon, giving confidence in the determination of the absolute cross sections in the present experiment performed in inverse kinematics.

\begin{figure}
\hspace*{-0.3cm}
\includegraphics[width=0.48\textwidth,clip]{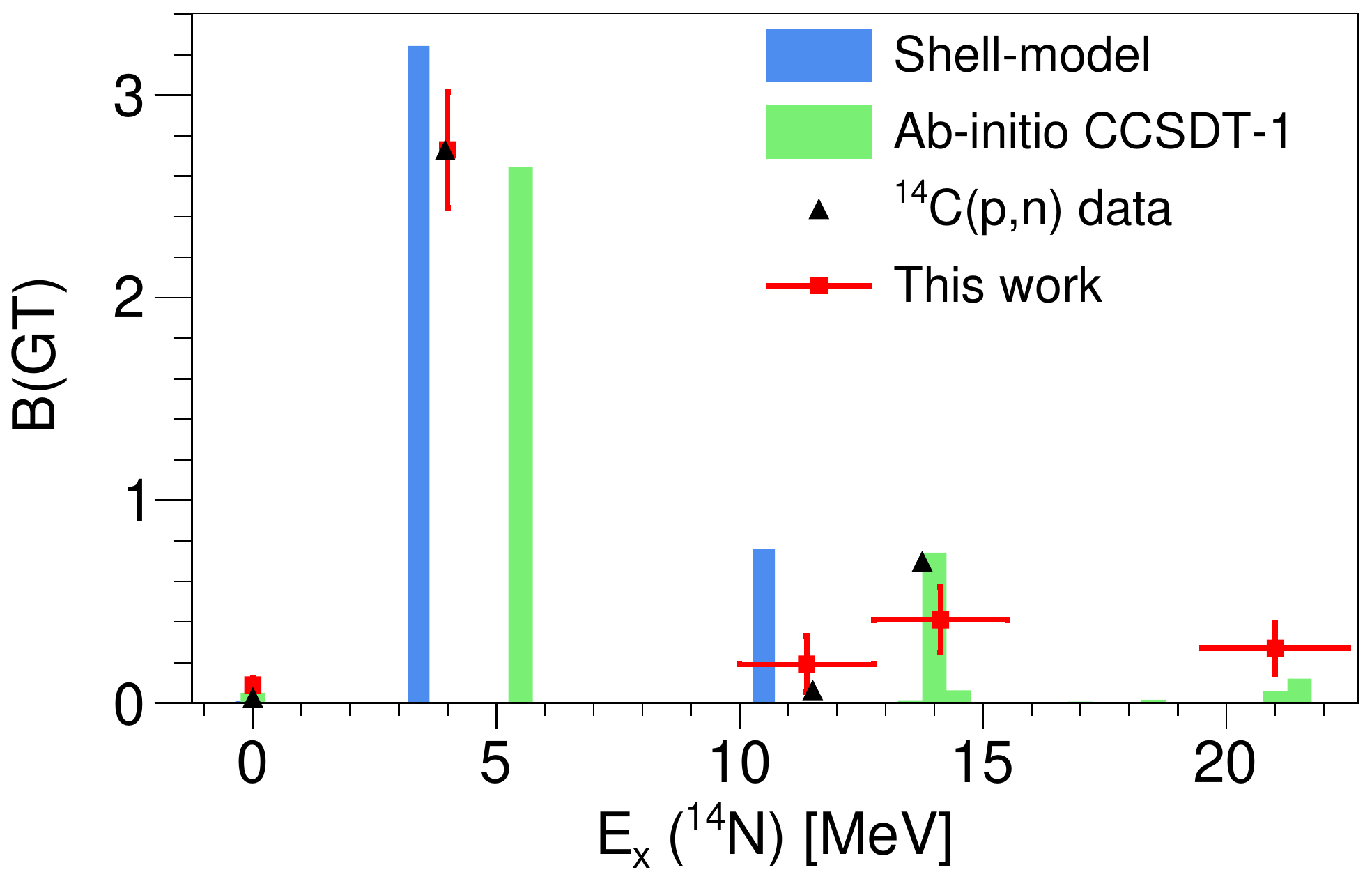}
\caption{ Comparison between the $B$(GT) distribution obtained from the present $^{14}$O$(d,{}^2\text{He})$ experiment, the analog $^{14}$C($p$,$n$) reaction at $E_{p}=160$ MeV~\cite{Goodman1980}, and theory by using SM and CC calculations (for details, see text). The horizontal error bars in the experimental data points from this work correspond to the regions in which the GT strength was observed and extracted from the MDA analysis.}\label{fig:bgt}
\end{figure}

Additional GT strength was identified in the excitation energy regions between 10 - 12.75 MeV (Fig.~\ref{fig:MDA}(c)), 12.75-15.5 MeV (Fig.~\ref{fig:MDA}(d)), and 19.5-22.5 MeV (Fig.~\ref{fig:MDA}(e)), and the associated $B$(GT)s were extracted by using the $\hat{\sigma}_{\textsc{GT}}$ obtain from the state at 3.95 MeV, as shown in Fig.~\ref{fig:bgt}. Other regions were also investigated, but no $\Delta L=0$ contributions were identified. As an example, the MDA for the region between 7-10 MeV is shown in Fig. \ref{fig:MDA}(f), indicating the dominance of transitions with $\Delta L>0$.   The GT strengths observed in the regions between 10 MeV - 12.75 MeV and 12.75-15.5 MeV most likely correspond to $1^{+}$ states observed at 11.5 MeV and 13.75 MeV in the isospin-symmetric $^{14}$C($p$,$n$) reaction ~\cite{Goodman1980,Taddeucci1984,Rapaport1987}, populating the same states in $^{14}$N. The summed strengths for these two states are very similar in the $(d,{}^2\text{He})$ and analog ($p$,$n$) experiments. The GT strength observed between 19.5 and 22.5 MeV was not observed in the $^{14}$C($p$,$n$) experiments. Unlike the latter experiments, in which contributions from the $^{12}$C($p$,$n$) reaction ($Q = -18.1$~MeV) due to $^{12}$C contaminants in the target made extraction of GT strength for the $^{14}$C($p$,$n$) reaction ($Q = -0.626$~MeV) difficult, the present result has no background and the higher-lying strength was unambiguously identified.

The transition to the $^{14}$N ground state is special. The known $B$(GT) of $2\times10^{-4}$ from ${}^{14}$O $\beta$ decay \cite{Hernandez1981} is well below the value for which the proportionality between $B$(GT) and the CE cross section holds \cite{Taddeucci1987,Zegers:2006}. The $^{14}$O$(d,{}^2\text{He})^{14}$N(g.s.) measured cross section associated with $\Delta L=0$ at $0^{\circ}$ is more than 100 times larger than expected based on the $B$(GT), which is comparable with the results from the $^{14}$N($^{3}$He,$t$)$^{14}$O(g.s.) \cite{PhysRevLett.97.062502} and the analog $^{14}$C($p$,$n$)$^{14}$N(g.s.) \cite{Taddeucci1987} reactions. To accurately describe the properties of this very weak transition, a consistent inclusion of 2- and 3-body (NN+3N) forces, and 2BCs is necessary \cite{PhysRevLett.113.262504}.

To test our understanding of the measured GT strength distribution, we compare with the SM calculations discussed above after applying a phenomenological quenching factor of 0.67 \cite{PhysRevC.47.163} and a coupled-cluster (CC) calculation in which the same NN+3N interaction and 2BCs of Ref.~\cite{Gysbers2019} are used. The effect of the inclusion of 2BCs amounts to a reduction in GT strength by a factor of 0.82. The CC calculation was performed using a natural orbital~\cite{Tichai2019} Hartree-Fock basis built from 15 major spherical oscillator shells with a frequency of $\hbar\omega=16$ MeV. We employed the chiral potential 1.8/2.0 (EM)~\cite{Hebeler2011} with 3N forces normal-ordered to the two-body level~\cite{Hagen2007}. With this basis, the non-Hermitian CC effective Hamiltonian was computed by solving the $^{14}$O ground state, which was used in turn to compute the corresponding left ground state~\cite{Bartlett2007,Hagen2014}. The Gamow-Teller response function was then computed using the equation-of-motion method for excited states~\cite{Stanton1993} and the Lanczos continued fraction method~\cite{Miroelli2016}. Both the ground and excited states were truncated at the singles, doubles, and approximate triples level (CCSDT-1)~\cite{Watts1995}. 

The comparisons between the experimental data and theory are shown in Fig. \ref{fig:bgt}. Overall, the theoretical calculations match the experimental data quite well. Aside from the transition to the ground state, the CC calculations put the strength at slightly higher excitation energy than the SM calculations, with the latter doing better for the strong transition to the 3.97-MeV state, and the CC calculation being more accurate for the strength between 10 and 15 MeV. In the present experiment, GT strength is found at $\sim21$ MeV. In contrast to the SM calculation, the CC calculations reproduce this strength, owing to the large-model space used. The summed experimental GT strength up to 22 MeV is $\Sigma B(\textrm{GT})=3.69 \pm0.75$, consistent with the CC calculations ($\Sigma B(\textrm{GT})=3.71$) and the SM calculations after quenching ($\Sigma B(\textrm{GT})=4.02)$. Clearly, with the newly developed $(d,{}^2\text{He})$ reaction in inverse kinematics, it is possible to test long-standing models such as the SM, as well as novel ab-initio calculations that include 3N forces and 2BCs. The good correspondence with the ab-initio calculations motivates further tests for nuclei with (very) asymmetric neutron-to-proton ratios.

In summary, we have demonstrated that the $(d,{}^2\text{He})$ reaction at $\sim100$ MeV/nucleon in inverse kinematics by using an active target time projection chamber placed in front of a magnetic spectrometer is an excellent method for model-independently extracting Gamow-Teller transitions strengths in the $\beta^{+}$ direction from unstable nuclei. We applied this method for the first time to extract the GT strength distribution from $^{14}$O and used it to test state-of-the-art CC calculations that take into consideration 3NF and 2BCs. In comparison with the SM, the CC calculations do not require a phenomenological quenching factor to reproduce the experimental strength distribution. By using the same experimental method similar detailed tests of theoretical models can be performed far from the valley of stability. This is not only of interest for testing ab-initio nuclear theories, but also to test a wider range of theoretical models, such as shell-models and density-functional theories, which are necessary for efficiently estimating GT transitions strength in the $\beta^{+}$/EC direction for a large number of nuclei. Such efforts will, for example, be important for estimating electron-capture rates in astrophysical scenarios. In combination with the previous development of the ($p$,$n$) reaction in inverse kinematics \cite{Sasano.Perdikakis.ea:2011}, experimental methods to use charge-exchange reactions to probe GT transition strengths from unstable nuclei beyond the $Q$-value window accessible through direct measurements of $\beta$ decay are now available in both $\beta^{+}$ and $\beta^{-}$ directions.

\begin{acknowledgments}
We thank all the staff at NSCL for their support. This work was supported by the US National Science Foundation under Grants PHY-1913554 (Windows on the Universe: Nuclear Astrophysics at the NSCL), PHY-1430152 (JINA Center for the Evolution of the Elements), PHY-2110365 (Nuclear Structure Theory and its Applications to Nuclear Properties, Astrophysics and Fundamental Physics).
AT-TPC was partially funded by the US National Science Foundation under Grant MRI-0923087. The work of S.~J.~N. was supported by the DOE Early Career Research Program, and G.~H. was supported by the Office of Nuclear Physics, U.S. Department of Energy, under grants DE-SC0018223 (NUCLEI SciDAC-4 collaboration) and by the Field Work Proposal ERKBP72 at Oak Ridge National Laboratory (ORNL). Computer time was provided by the Innovative and Novel Computational Impact on Theory and Experiment (INCITE) program, and used resources at ORNL which is supported by the Office of Science of the Department of Energy under Contract No. DE-AC05-00OR22725. The work of S.~B. was supported by the Deutsche Forschungsgemeinschaft through the Cluster of Excellence "Precision Physics, Fundamental Interactions, and Structure of Matter" (PRISMA$^+$ EXC 2118/1, Project ID 39083149).
\end{acknowledgments}

%\end{linenumbers}

% Create the reference section using BibTeX
\bibliography{biblio}

\end{document}